\documentclass[12pt,draftcls,onecolumn]{IEEEtran}

\newtheorem{TT}{Theorem}
\newtheorem{CC}{Corollary}

\ifCLASSINFOpdf
\else
\fi

\ifCLASSOPTIONcompsoc
\usepackage[nocompress]{cite}
\else
\usepackage{cite}
\fi

\usepackage{amsmath}
\usepackage{amssymb}
\usepackage{graphicx}
\usepackage{epstopdf}
\usepackage{mathtools}
\usepackage{xcolor}

\begin{document}

\title{On  BT-limited Signals}


\author{
        Xiang-Gen Xia, \IEEEmembership{Fellow}, \IEEEmembership{IEEE} 

\thanks{
X.-G. Xia is with the Department of Electrical and Computer Engineering, University of Delaware, Newark, DE 19716, USA (e-mail: xxia@ece.udel.edu).}

}

\date{}

\maketitle


\begin{abstract}
In this paper, we introduce and characterize 
a subspace of  bandlimited signals.
The subspace consists of all $\Omega$ bandlimited signals 
such that the non-zero parts of their Fourier transforms
are pieces of some $T$ bandlimited signals. 
The signals in the subspace  are called 
{\em BT-limited signals} and 
the subspace is named as  BT-limited signal space. 
For BT-limited signals, 
a signal extrapolation with an analytic error estimate exists
outside the interval $[-T, T]$ of given signal values with errors. 
Some new properties about and applying BT-limited signals are also presented. 
\end{abstract}



\section{Introduction}

Bandlimited signals have played a fundamental role in the digital world 
in the last decades. The Whittaker-Shannon sampling theorem is the fundamental 
bridge between analog and digital signal processings/communications,
 which has brought 
significant interest in both signal processing and mathematics communities. 
The sampling theorem is about the reconstruction of a bandlimited signal 
from its evenly spaced samples and an exact 
reconstruction is possible if the samples are sampled 
from the analog signal with a sampling rate not lower than the Nyquist rate. 

Another family of bandlimited signal reconstructions is to reconstuct a 
bandlimited signal from its given segment.  It is called 
bandlimited signal extrapolation and has applications in, for example, 
CT imaging, where only limited observation angles are available. 
Bandlimited signal extrapolation has also attracted significant 
interest in the past, see, for example, \cite{gerch}-\cite{xia}. 

In this paper, we introduce a subspace of bandlimited signals, which 
is called {\em BT-limited} signal space. It consists of all bandlimited 
signals such that the non-zero parts of their Fourier transforms
are pieces of bandlimited signals, which are called BT-limited signals. 
Note that for a general bandlimited signal, although its Fourier 
transform has finite support, the non-zero spectrum  may not be smooth,
while the non-zero spectrum is smooth for a BT-limited signal. 
It was found in \cite{xia} that BT-limited signals 
can be characterized by using prolate spheroidal wavefunctions \cite{slepia}.
In this paper, a more intuitive and elementary 
proof for the characterization  is given,
which may help to better understand BT-limited signals. 
Some new properties about and applying BT-limited signals are also presented. 
Interestingly, although there is no any error estimate existed 
for a general bandlimited signal extrapolation from inaccurate data,
an analytic error estimate in the whole time domain was obtained 
in \cite{xia} for a BT-limited signal extrapolation. 

The remainder of this paper is organized as follows. 
In Section \ref{sec2},  prolate
spheroidal wavefunctions are briefly introduced. 
In Section \ref{sec3}, BT-limited signals are introduced and characterized.
In Section \ref{sec4}, a BT-limited signal extrapolation with analytic error
estimate is described. In Section \ref{sec5}, some simulations 
are presented to verify the theoretical 
extrapolation result for BT-limited signals. 
In Section \ref{sec6}, more properties on BT-limited signals are presented.
In Section \ref{sec7}, this paper is concluded. 

\section{Bandlimited Signal Space}\label{sec2}

All signals considered in this paper  are assumed to have finite energies. 
A signal $f(t)$ is called bandlimited of bandwidth $\Omega$ (or $\Omega$
bandlimited),  if its Fourier 
transform $\hat{f}(\omega)$ vanishes when $|\omega|>\Omega$. 
Let ${\cal BL}_{\Omega}$ denote the space of 
all $\Omega$ bandlimited signals.
Let $T>0$ be a constant and $K$ be the following operator defined on
$L^2[-T,T]$:
\begin{equation}\label{1}
(Kf)(t)=\int_{-T}^{T} \frac{\sin \Omega (t-s)}{\pi (t-s)} f(s) ds,
\,\,\,\mbox{ for } f\in L^2[-T,T].
\end{equation}
Let $\phi_k$ and $\lambda_k$, $k=0,1,2,...$, be the eigenfunctions 
and the corresponding eigenvalues of the operator $K$ with
\begin{equation}\label{100}
\int_{-T}^{T} \phi_j(t)\phi_k(t)dt = \lambda_k \delta (j-k),
\end{equation}
where $\delta(n)$ is $1$ when $n=0$ and $0$ otherwise, 
and $1>\lambda_0> \lambda_1>\cdots>0$ with $\lambda_k \rightarrow 0$ as
$k\rightarrow \infty$.

From (\ref{1}), for $k=0,1,2,...$, 
\begin{equation}\label{ext}
\phi_k(t)= \frac{1}{\lambda_k}
\int_{-T}^{T} \frac{\sin \Omega (t-\tau)}{\pi (t-\tau)} \phi_k(\tau) d\tau,
\,\,\,\mbox{ for } t\in [-T,T],
\end{equation}
which means that $\phi_k(t)$ can be extended from $t\in [-T, T]$ to
$t\in (-\infty, \infty)$. Then,
$$
\int_{-\infty}^{\infty} \phi_j(t) \phi_k (t)dt =\delta(j-k)
$$
and $\{\phi_k(t)\}_{k=0}^{\infty}$ 
form an orthonormal basis for space ${\cal BL}_{\Omega}$
and every $\Omega$ bandlimited signal $f$ can be expanded as
\begin{equation}\label{2}
f(t)=\sum_{k=0}^{\infty} a_k \phi_k(t)
\end{equation}
for some constants $a_k$ with
$$
\sum_{k=0}^{\infty}|a_k|^2 =\| f\|^2 <\infty.
$$
The extended eigenfunctions $\phi_k$ are called 
{\em  prolate spheroidal wavefunctions} \cite{slepia}.

\section{BT-limited Signal Space}\label{sec3}

We next define a subspace of $\Omega$ bandlimited signals.
An $\Omega$ bandlimited signal is called {\em BT-limited} if 
the non-zero part of its Fourier transform is a piece of a  
$T$ bandlimited signal. In other words, let  $f\in {\cal BL}_{\Omega}$
and its Fourier transform be $\hat{f}$. If there exists $g\in {\cal BL}_T$
and $\hat{f}(\omega)=g(\omega)$ for $\omega\in [-\Omega, \Omega]$, then $f$ 
is called BT-limited. The subspace of all  BT-limited 
signals in  ${\cal BL}_{\Omega}$ is denoted as ${\cal BL}_{\Omega}^0$
and called {\em BT-limited signal space}. 

From the above definition, by taking 
Fourier transform and  inverse Fourier transfrom,
 it is not hard to see that 
 $f(t)\in {\cal BL}_{\Omega}^0$
if and only if there exists $q(t)\in L^2[-T,T]$ such that
\begin{equation}\label{400}
f(t)= \int_{-T}^{T} \frac{\sin \Omega (t-s)}{\pi(t-s)} q(s) ds, \,\, \mbox{ for } t\in (-\infty, \infty). 
\end{equation}
Thus, from (\ref{ext}), we know that every prolate spheroidal wavefunction
$\phi_k$ is BT-limited, $\phi_k\in {\cal BL}_{\Omega}^0$, so is any linear 
combination of finite many prolate spheroidal wavefunctions.

For a general $\Omega$ bandlimited signal, although its Fourier transform
has  finite suppport, the non-zero part of the Fourier transform 
is only in $L^2[-\Omega, \Omega]$ and may not be smooth. However, a BT-limited signal is not only smooth 
(entire function of exponential type \cite{boas}) 
in time domain but also has the same smoothness for the non-zero 
part in frequency domain. 
To characterize ${\cal BL}_{\Omega}^0$, the following result was obtained  in \cite{xia}. 

\begin{TT} 
Let $f\in {\cal BL}_{\Omega}$ with the expansion (\ref{2}).
Then, $f\in {\cal BL}_{\Omega}^0$ if and only if 
\begin{equation}\label{33}
\sum_{k=0}^{\infty}
\frac{|a_k|^2}{\lambda_k} <\infty.
\end{equation}
\end{TT}

The proof given in \cite{xia} is based on a result on operator theory. 
Below, we provide an elementary and intuitive proof of Theorem 1,
which may help to understand BT-limited signals better. 

\noindent
{\bf Proof}:

We first prove the ``if'' part. 
Let $f(t)$ be an $\Omega$ bandlimited signal with the expansion (\ref{2}) and 
the property (\ref{33}) hold. Let
\begin{equation}\label{proof1}
q(t)=\sum_{k=0}^{\infty} \frac{a_k}{\lambda_k} \phi_k(t). 
\end{equation}
From (\ref{100}) and (\ref{33}), we have
$$
\int_{-T}^T |q(t)|^2 dt = \sum_{k=0}^{\infty}\frac{|a|_k^2}{\lambda_k}<\infty.
$$
Thus, $q(t)\in L^2[-T, T]$. Furthermore, from (\ref{1}), we have 
$$
(Kq)(t)=\sum_{k=0}^{\infty} a_k \phi_k(t) =f(t), \,\,\mbox{ for }
t\in (-\infty, \infty),
$$
which is (\ref{400}) and therefore, $f(t)\in {\cal BL}_{\Omega}^0$.
This proves the sufficiency.

We next prove the ``only if'' part. 
If $f(t)\in {\cal BL}_{\Omega}^0$, 
then $f(t)$ has the form (\ref{400}) for some $q(t)\in L^2[-T, T]$. 
In the meantime, since $f(t)$ is $\Omega$ bandlimited, let 
$f(t)$ have the expansion (\ref{2}). 
Since $\{\phi_k(t)\}_{k=0}^{\infty}$ form an orthogonal basis for 
$L^2[-T, T]$, \cite{slepia}, and (\ref{100}), there exist a sequence of constants
$\{b_k\}_{k=0}^{\infty}$ of finite energy, i.e., 
\begin{equation}\label{proof2}
\sum_{k=0}^{\infty} |b_k|^2<\infty,
\end{equation}
such that
$$
q(t)=\sum_{k=0}^{\infty} b_k \frac{\phi_k(t)}{\sqrt{\lambda_k}}, \,\,
\mbox{ for } t\in [-T, T].
$$
Thus, from (\ref{400}) and (\ref{1}) , we have 
$$
f(t)=(Kq)(t)=\sum_{k=0}^{\infty} b_k \sqrt{\lambda_k} \phi_k(t),\,\,
\mbox{ for } t\in (-\infty, \infty).
$$
Therefore, comparing with (\ref{2}), we obtain $a_k=b_k\sqrt{\lambda_k}$
for $k=0,1,2,...$. From (\ref{proof2}), we then have 
$$
\sum_{k=0}^{\infty} \frac{|a_k|^2}{\lambda_k}
=\sum_{k=0}^{\infty} |b_k|^2<\infty,
$$
which proves (\ref{33}), i.e., the necessity is proved. 
{\bf q.e.d.}

The above result characterizes all BT-limited signals.
Since any linear combinations of finite many prolate spheroidal wavefunctions 
are BT-limited, all BT-limited signals are dense 
in a bandlimited signal space, i.e., any bandlimited signal can be 
approximated by BT-limited signals. 

Since $\{\phi_k\}_{k=0}^{\infty}$ form an orthonormal basis for 
space ${\cal BL}_{\Omega}$, \cite{slepia}, for any finite energy sequence $\{a_k\}_{k=0}^{\infty}$, i.e.,
$$
\sum_{k=0}^{\infty} |a_k|^2<\infty,
$$
we know 
$$
\sum_{k=0}^{\infty}a_k \phi_k(t)\in {\cal BL}_{\Omega}.
$$
On the other hand, from (\ref{100}), 
$$
\sum_{k=0}^{\infty}a_k \frac{\phi_k(t)}{\sqrt{\lambda_k}} 
=\sum_{k=0}^{\infty}\frac{a_k}{\sqrt{\lambda_k}} \phi_k(t)
\in L^2[-T,T].
$$
Since $\lambda_k\rightarrow 0$ as $k\rightarrow \infty$,
we may have
$$
\sum_{k=0}^{\infty}\frac{|a_k|^2}{\lambda_k}=\infty.
$$
Thus, in general
$$
\sum_{k=0}^{\infty}a_k \frac{\phi_k(t)}{\sqrt{\lambda_k}} \notin 
L^2(-\infty, \infty),\,\,
\mbox{ or equivalently, }
\notin {\cal BL}_{\Omega}.
$$
However, if 
$$
\sum_{k=0}^{\infty}a_k \frac{\phi_k(t)}{\sqrt{\lambda_k}} \in 
L^2(-\infty, \infty),\,\,
\mbox{ or equivalently, }
\in {\cal BL}_{\Omega},
$$
then, (\ref{33}) holds, and from  Theorem 1, we obtain 
$$
\sum_{k=0}^{\infty}a_k \frac{\phi_k(t)}{\sqrt{\lambda_k}} 
\in {\cal BL}_{\Omega}^0. 
$$

For a general $\Omega$ bandlimited signal $f\in {\cal BL}_{\Omega}$
with expansion (\ref{2}) where $\{a_k\}_{k=0}^{\infty}$ is a general 
sequence of finite energy, let 
\begin{equation}\label{fn}
f_n(t)=\sum_{k=0}^{n}a_k\phi_k(t).
\end{equation}
From (\ref{1}) and (\ref{ext}), 
$$
f_n(t)= K \left( \sum_{k=0}^{n} \frac{a_k}{\lambda_k}\phi_k(t)\right).
$$
Let 
$$
q_n(t)=\sum_{k=0}^{n} \frac{a_k}{\lambda_k}\phi_k(t).
$$
Clearly, we have $f_n(t)=(Kq_n)(t)\in {\cal BL}_{\Omega}^0$.
However, 
 if $f$ is not BT-limited,
i.e., $f\notin {\cal BL}_{\Omega}^0$, then, from (\ref{100}) and Theorem 1, 
$$
\int_{-T}^T |q_n(t)|^2dt 
=\sum_{k=0}^n \frac{|a_k|^2}{\lambda_k} \rightarrow \infty,\,\,\mbox{ as }
n\rightarrow \infty.
$$
This means that $q_n(t)$ does not converge in $L^2[-T,T]$,
although $f_n(t)$ converges to $f(t)$ in $L^2(-\infty,\infty)$,
 as $n\rightarrow \infty$, otherwise $f$ would be BT-limited.

Since $\Omega$ bandlimited signal space ${\cal BL}_{\Omega}$ and 
$L^2[-\Omega, \Omega]$ are isomorphic  by using (inverse) Fourier 
transform, for any signal $g\in L^2[-\Omega, \Omega]$, let it 
be the Fourier transform $\hat{f}$ of $f\in {\cal BL}_{\Omega}$, i.e.,
$g=\hat{f}$ on $[-\Omega, \Omega]$. 
As we can see above, $f_n$ approaches $f$ in 
$L^2(-\infty, \infty)$, then $\hat{f}_n$ approaches $g=\hat{f}$
in $L^2[-\Omega, \Omega]$. Since $\hat{f}_n= D_{\Omega} \hat{q}_n$
and $\hat{q}_n$ is $T$ bandlimited as we can see above as well, 
$g$ can be approximated by a $T$ bandlimited signal $\hat{q}_n$
restricted in $[-\Omega, \Omega]$ 
in $L^2[-\Omega, \Omega]$, where $D_{\Omega}$ stands for the 
truncation operator from $(-\infty, \infty)$ to $[-\Omega, \Omega]$. 
Because $T$ and $\Omega$ are both arbitrary, the above
analysis proves the following corollary.

\begin{CC}
Any finite piece signal on $[a, b]$ can be approximated 
in $L^2[a, b]$  by a bandlimited signal restricted 
in $[a, b]$ of  bandwidth $T$,
where $-\infty <a<b<\infty$ and $T>0$ are arbitrary. 
\end{CC}

Note that the above result does not hold for  infinite length signals.
Also, as a comparison, 
 the Weierstrass theorem says that any finite piece continuous 
signal can be approximated by polynomials,
which may be thought of as  a different perspective 
of using smooth/simple signals to approximate complicated signals.

\section{BT-limited Signal Extrapolation}\label{sec4}

Bandlimited signal extrapolation had been studied extensively in the 1970s and 1980s, see, for example, \cite{gerch}-\cite{xia}. 
It is to extrapolate a bandlimited signal $f$ from a given piece of its values,
for example, to extrapolate $f(t)$ for $t$ outside $[-T, T]$ when 
$f(t)$ for $t\in [-T, T]$ is given. It is possible in theory since 
$f$ is bandlimited and thus it is an entire function \cite{boas}. Any entire function is 
completely determined by its any segment. However, in practice, 
a given piece signal $f(t)$ for $t\in [-T, T]$ may contain error/noise and in this case, 
the extrapolation problem becomes a well-known ill-posed inverse problem. 
Any error in a given segment may cause an arbitrary large error in an extrapolation 
in general. 

However, when $f$ is BT-limited, an extrapolation method was proposed in \cite{xia}
and an analytic error estimate for the extrapolation over the whole time domain
was obtained. It can be described as follows. 


Let $f_{\epsilon}(t)$ be an observation of $f(t)$ for $t\in [-T, T]$ with 
the maximal error magnitude $\epsilon$, i.e., 
$|f_{\epsilon}(t)-f(t)|\leq \epsilon$ for 
$t\in [-T, T]$. Let $q_{\epsilon}$ be the following minimum norm solution 
(MNS) in space
$L^2[-T, T]$:
\begin{eqnarray}
\int_{-T}^{T} |q_{\epsilon}(t)|^2dt 
 & = & \min_{q(t)\in L^2[-T, T]} \left\{ \int_{-T}^{T} |q(t)|^2dt\, : \right.
  \nonumber \\
 &   & \left. \left| \int_{-T}^{T} \frac{\sin \Omega (t-s)}{\pi(t-s)} q(s) ds -f_{\epsilon}(t)
\right|\leq 2\epsilon, \mbox{ for } t\in [-T, T] 
\right\}.  \label{3}
\end{eqnarray}
Let 
\begin{equation}\label{4}
\tilde{f}(t)= \int_{-T}^{T} \frac{\sin \Omega (t-s)}{\pi(t-s)} q_{\epsilon}(s) ds.
\end{equation}
One can see that the above $\tilde{f}(t)$ is obtained from the given observation 
segement $f_{\epsilon} (t)$ of $f(t)$ on $[-T, T]$ and is called an extrapolation
of $f(t)$. 
Also,  from (\ref{400}),  we have $\tilde{f}(t)\in {\cal BL}_{\Omega}^0$.
For the above extrapolation of $f(t)$, 
the following result was obtained in \cite{xia}.

\begin{TT} 
If $f$ is BT-limited, i.e.,  $f\in {\cal BL}_{\Omega}^0$, and $\tilde{f}$ is defined 
in (\ref{4}), then 
\begin{equation}\label{5}
|\tilde{f}(t)-f(t)|\leq C \epsilon^{1/3},\,\,\, \mbox{ for all }
t\in (-\infty, \infty),
\end{equation}
for some constant $C$ that is independent of $\epsilon$ and $t$. 
\end{TT}

This result tells that when signal $f$ is BT-limited, i.e., 
not only it is bandlimited but also the non-zero part of its Fourier 
transform is  a piece of a bandlimited signal, the above extrapolation
(\ref{3})-(\ref{4}) is robust and has an error estimate (\ref{5}) for time $t$. To the author's best knowledge, no any other error estimate 
for a bandlimited signal extrapolation from inaccurate data 
 on the whole time domain exists 
in the literature. 

In practice, a given observation $f_{\epsilon}(t)$ for $t\in [-T, T]$ is usually 
discrete in time. A discretization of the above extrapolation (\ref{3})-(\ref{4}) 
with a proved convergence was also given in \cite{xia}.

More general subspaces ${\cal BL}_{\Omega}^{\gamma}$ 
for $0\leq \gamma <1/2$ in bandlimited signal space ${\cal BL}_{\Omega}$
 than the above ${\cal BL}_{\Omega}^0$ were introduced with the corresponding
extrapolation, error estimate and discretization in \cite{xia}. 
It was shown in \cite{xia} that,  if $f(t)$
is $\Omega$ bandlimited with the expansion (\ref{2}) and, for $0\leq \gamma<1/2$, the following inequality holds 
\begin{equation}\label{333}
\sum_{k=0}^{\infty}
\frac{|a_k|^2}{\lambda_k^{1-2\gamma/3}} <\infty,
\end{equation}
then, $f(t)\in {\cal BL}_{\Omega}^{\gamma}$. Clearly, (\ref{333}) 
 returns to (\ref{33}) in 
Theorem 1 when $\gamma=0$, although when $\gamma\neq 0$, 
the physical meaning of signals in subspace ${\cal BL}_{\Omega}^{\gamma}$
is not as clear as signals in subspace  ${\cal BL}_{\Omega}^{0}$ studied in this paper. 
For more details, we refer the reader to \cite{xia}. 

Another comment we want to make here is that, for a general 
$\Omega$ bandlimited signal $f$, although it may not be 
BT-limited, function $f_n$ defined in (\ref{fn}) 
is BT-limited and  approaches $f$ in $L^2(-\infty, \infty)$
 as $n$ becomes large. 
Therefore, for a general $\Omega$ bandlimited signal $f$, 
from its given segment 
$\tilde{f}$ with errors, we can still apply the MNS 
extrapolation (\ref{3})-(\ref{4}). In this case, 
 the analytic error estimate in Theorem 2 may not hold. However, 
interestingly, it was shown in \cite{xia}
that the discretization of the above MNS extrapolation
and its convergence to the analog solution 
still hold.

\section{Simulations}\label{sec5}
We next show some simulation results to verify the above 
MNS extrapolation for BT-limited signals. 
For simplicity, in this simulation 
we use $\Omega=\pi$ and $T=1$. A BT-limited signal
$f(t)$ is generated by randomly generating $q(t)\in L^2[-T, T]$ in 
(\ref{400}). Its noisy observation $f_{\epsilon}(t)$ is obtained by
adding a random error with uniform distribution to $f(t)$ so that the maximum error
magnitude not above $\epsilon$. 

\begin{figure}
\centering
\includegraphics[scale=0.6]{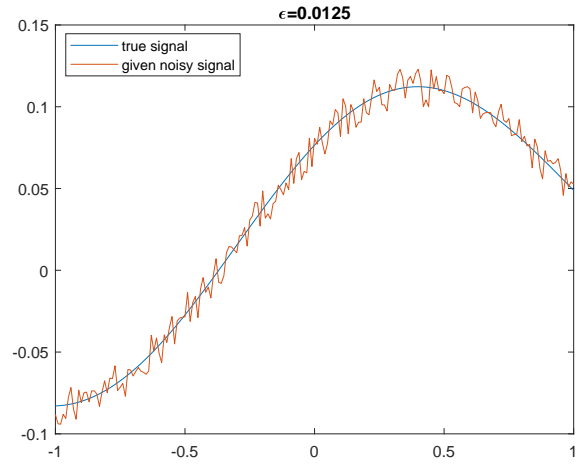}\\
(a)\\
\includegraphics[scale=0.6]{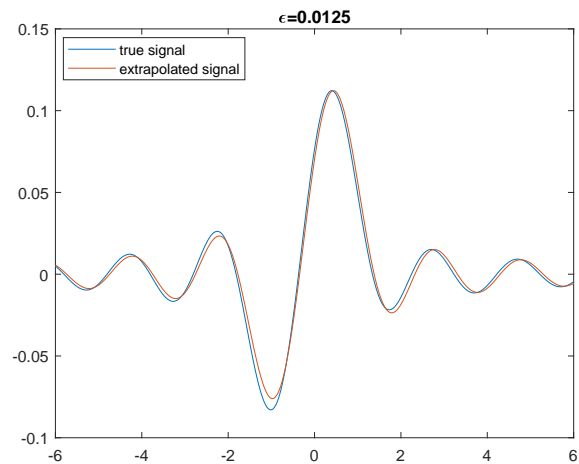}\\
 (b)
\caption{BT-limited signal extrapolation from noisy data with the maximum 
error magnitude $\epsilon=0.0125$: (a) given noisy
data on $[-1, 1]$  and (b) extrapolated signal using the MNS method.}
\label{fig1}
\end{figure}

\begin{figure}
\centering
\includegraphics[scale=0.6]{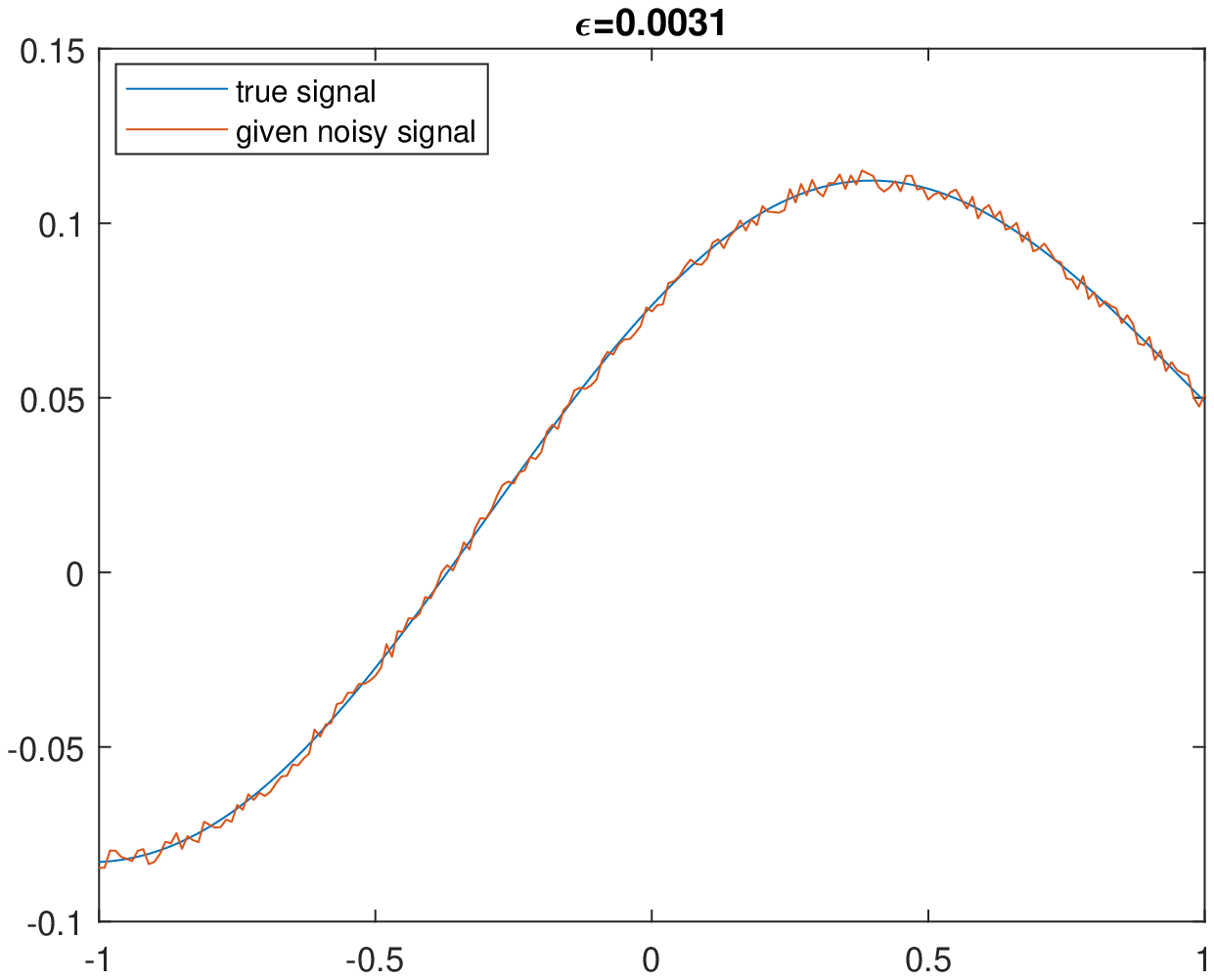}\\
(a)\\
\includegraphics[scale=0.6]{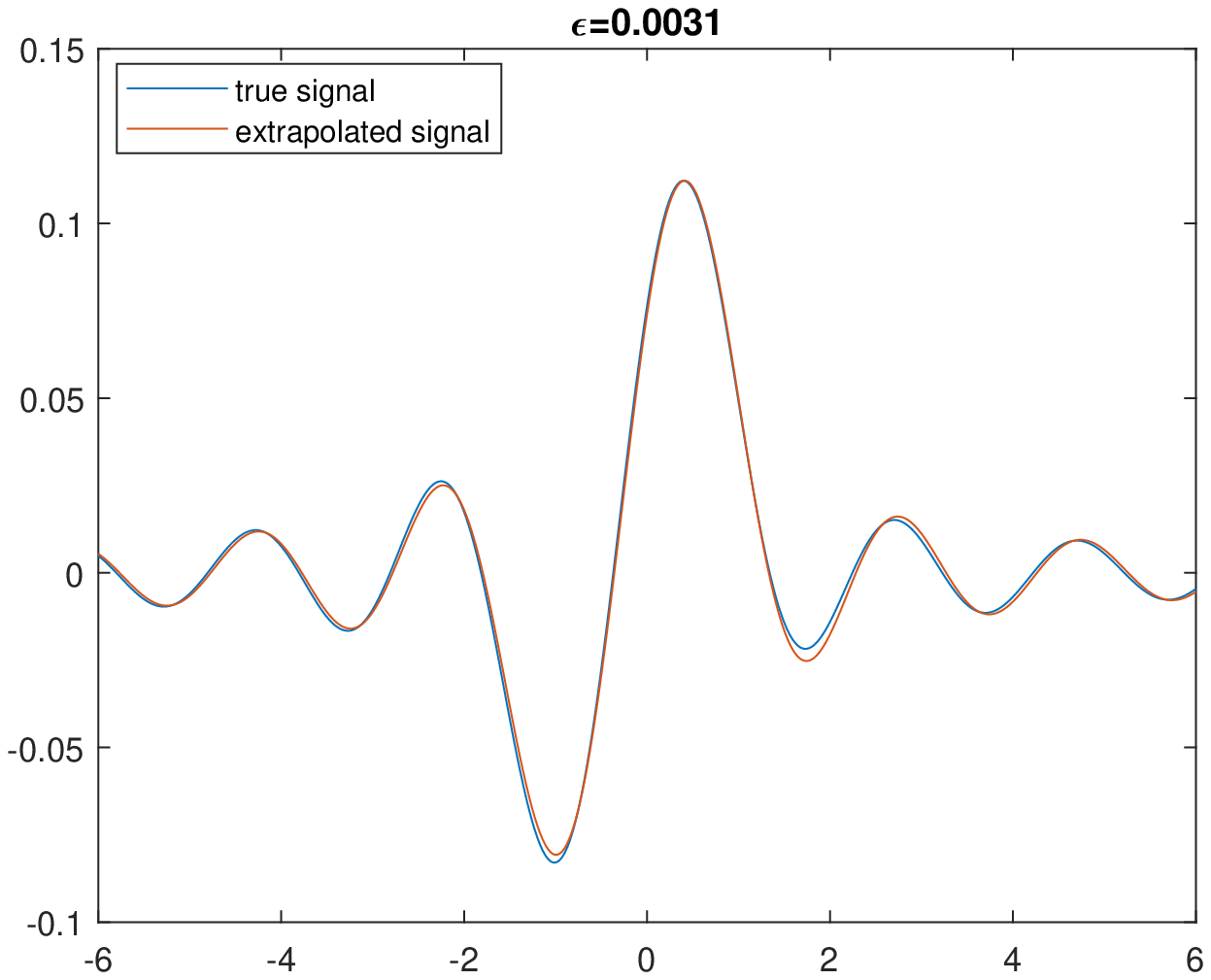}\\
(b)

\caption{BT-limited signal extrapolation from noisy data with the maximum 
error magnitude $\epsilon=0.0031$: (a) given noisy
data on $[-1, 1]$  and (b) extrapolated signal using the MNS method.}
\label{fig2}
\end{figure}

We sample a noisy analog BT-limited signal $f_{\epsilon}(t)$ in $[-1, 1]$ with sampling rate $100$ Hz, i.e., $201$ samples of $f_{\epsilon}(t)$
 in $[-1,1]$ are used in the MNS in (\ref{3})-(\ref{4}).
In Fig. \ref{fig1}, the case of $\epsilon=0.0125$ is simulated, where 
Fig. \ref{1}(a) shows the true data of a BT-limited signal $f(t)$ 
and its noisy data $f_{\epsilon}(t)$ on $[-1,1]$, and Fig. \ref{fig1}(b)
shows the true signal $f(t)$ and its extrapolation $\tilde{f}(t)$
in (\ref{4}) using the noisy  data $f_{\epsilon}(t)$ shown in 
Fig. \ref{1}(a). 
Fig. \ref{fig2} shows the results when $\epsilon=0.0031$, where one can 
 see that the error in the extrapolated signal is clearly reduced, 
comparing to that  in Fig. \ref{fig1}.

\begin{figure}[t]
\centering

\includegraphics[scale=0.6]{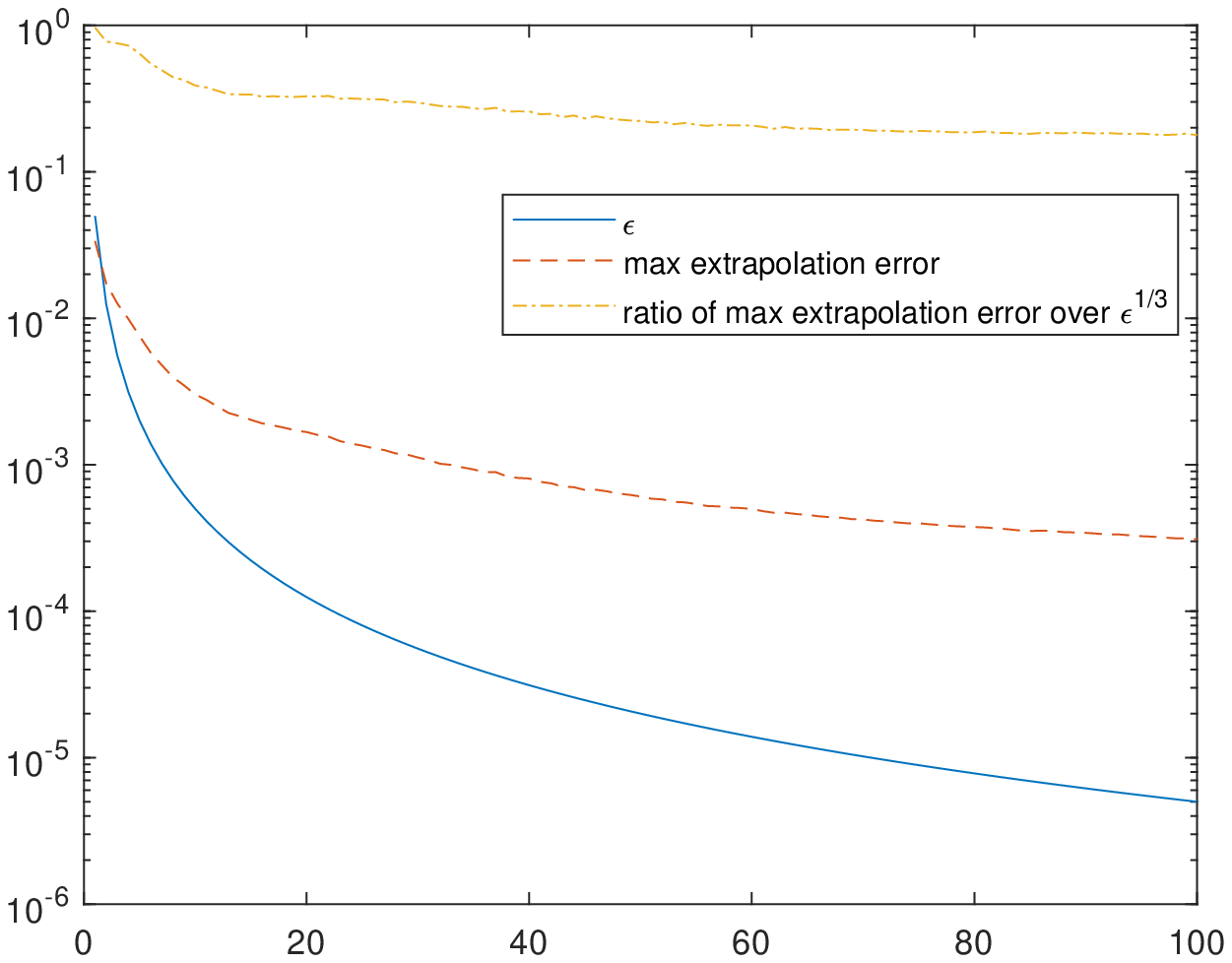}

\caption{The maximum error between extrapolated and true signals vs. 
the maximum magnitude $\epsilon$ of the errors in the given data over the 
time interval $[-1,1]$, and its ratio, $R(\epsilon)$, over $\epsilon^{1/3}$.}

\label{fig3}
\end{figure}

Fig. \ref{fig3} shows the curve (dashed)
 of the  maximum error magnitude between
the true and the extrapolated signals, i.e.,
$\max_t |f(t)-\tilde{f}(t)|$,  vs. the maximum error magnitude
$\epsilon$ in the noisy data over $[-1,1]$, and the 
curve (dashdot) of the ratio vs. $\epsilon$: 
$$
R(\epsilon)= \frac{\max_t |f(t)-\tilde{f}(t)|}{\epsilon^{1/3}}.
$$ 
The curves are obtained by using $20$ independent
trials. 
From this figure, one can see that the ratio
$R(\epsilon)$ is less than a constant as
$\epsilon$ gets smaller, which verifies the result (\ref{5}) in Theorem 2. 
Note that in Fig. \ref{fig3}, 
the signal magnitudes are similar to those in Figs. \ref{fig1} 
and \ref{fig2}.

\section{More Properties of BT-limited Signals}\label{sec6}
The above definition of a BT-limited signal can be easily generalized
as follows. Let ${\cal BL}_{[A, B]}$ denote 
the space of all finite energy signals $f(t)$ whose Fourier transforms
are supported in the interval $[A, B]$, i.e., 
$\hat{f}(\omega)=0$ when $\omega\notin [A, B]$. 
For real numbers $A, B, a, b$ with $-\infty<A<B<\infty$ and $-\infty <a<b<\infty$, 
if signal $f(t)\in {\cal BL}_{[A, B]}$ 
and $\hat{f}(\omega)=g(\omega)$ when 
$\omega\in [A, B]$ for some 
$g(\omega) \in {\cal BL}_{[a, b]}$, 
then signal $f(t)$ is called BT-limited.
The signal space of
all the above BT-limited signals is denoted as ${\cal BL}_{A,B,a,b}$. 
Clearly, when $A=-\Omega$, $B=\Omega$, $a=-T$, and $b=T$, the above definition
for a BT-limited signal 
returns to that in Section \ref{sec3} and ${\cal BL}_{A,B,a,b}={\cal BL}_{\Omega}^0$.

Let $\Omega=(B-A)/2$ and $T=(b-a)/2$, by some shifts in frequency and time 
domains, the representation for a BT-limited signal in 
 (\ref{400}) becomes as follows: 
$f\in {\cal BL}_{A,B,a,b}$ if and only if 
\begin{equation}\label{61}
f(t)= e^{j (A+\Omega)t}
\int_{-T}^{T} \frac{\sin \Omega (t+a+T-s)}{\pi(t+a+T-s)} q(s) ds, \,\, \mbox{ for } t\in (-\infty, \infty),
\end{equation}
for some $q(t)\in L^2[-T,T]$.

Since any bandlimited signal is an entire function when $t$ is extended 
to the complex plane \cite{boas}, it cannot be $0$ in any segment 
of time domain unless it is all $0$ valued. This implies that
a bandlimited signal $f$ whose Fourier transform is supported in
two separate bands, for example, $\hat{f}(\omega)\neq 0$ 
for $A_i<\omega<B_i$, $i=1,2$,  and $\hat{f}(\omega)=0$ for other $\omega$,
where $-\infty<A_1<B_1<A_2<B_2<\infty$, then, signal $f$ is not BT-limited,
i.e., $f\notin {\cal BL}_{A_1, B_2, a, b}$ for any $-\infty<a<b<\infty$, although
in this case, signal $f$ could be  a sum of two BT-limited signals
whose Fourier transforms are supported in $[A_1, B_1]$ and $[A_2, B_2]$,
respectively, 
such that,  the non-zero supports of the two Fourier transforms 
are the pieces of two bandlimited signals. 

\begin{TT}\label{thm3}
For two non-zero BT-limited signals $f_i\in {\cal BL}_{A_i, B_i, a_i, b_i}$
with $-\infty <A_i<B_i<\infty$ and $-\infty <a_i<b_i<\infty$
 for $i=1,2$, their linear 
combination $f= \alpha_1 f_1 + \alpha_2 f_2$ with two non-zero complex 
coefficients $\alpha_1$ and $\alpha_2$  is  BT-limited {\em if and only if}
$A_1=A_2$ and $B_1=B_2$.  
\end{TT}

\noindent
{\bf Proof}:

The ``if'' part is easy to see by setting
$A=A_1=A_2$, $B=B_1=B_2$, $a=\min\{a_1, a_2\}$, and $b=\max\{b_1,b_2\}$.
Then, $f\in {\cal BL}_{A,B,a,b}$, i.e., $f$ is BT-limited.

We next prove the ``only if'' part. 
Without loss of generality, 
we assume  $A_1<A_2$  and $f$ is BT-limited.  Then, $f\in {\cal BL}_{A_1, 
\max\{B_1, B_2\}, a, b}$ for some real numbers 
$a, b$ with $-\infty <a<b<\infty$.
From the above 
definition of BT-limited signals, there exist bandlimited signals 
$g_i\in {\cal BL}_{a_i, b_i}$ such that $\hat{f}_i(\omega)=g_i(\omega)$
for $\omega\in [A_i, B_i]$,   $i=1,2$, 
and there exists a bandlimited signal $g\in {\cal BL}_{a,b}$ such that
$\hat{f}(\omega)=g(\omega)$ for $\omega\in [A_1, \max\{B_1, B_2\}]$.
On the other hand, we have 
$\hat{f}=\alpha_1 \hat{f}_1 +\alpha_2 \hat{f}_2$. 
This means that 
$\hat{f}(\omega)=\alpha_1 \hat{f}_1(\omega)=g(\omega)=\alpha_1 g_1(\omega)$ for
 $\omega\in [A_1, \min\{A_2, B_1\}]$. Since all
$g, g_1, g_2$ are bandlimited and therefore, entire functions, 
we must have
$\hat{f}(\omega)=\alpha_1 \hat{f}_1(\omega)=g(\omega)=\alpha_1 g_1(\omega)$ for
all $\omega$. In other words, 
$\hat{f}(\omega)=\alpha_1 \hat{f}_1(\omega)+\alpha_2 \hat{f}_2(\omega)
=\alpha_1 \hat{f}_1(\omega)$ for all $\omega$.
This implies $\hat{f}_2(\omega)=0$ for all $\omega$, i.e., 
$f_2$ is the $0$ signal,  which contradicts 
with the non-zero signal assumption. 
This proves the necessity. 
 {\bf q.e.d.}

In general, for $p$ BT-limited signals $f_i\in {\cal BL}_{A_i, B_i, a_i, b_i}$,
with $-\infty <A_i<B_i<\infty$ and $-\infty <a_i<b_i<\infty$,  $i=1,2,...,p$, 
their non-zero linear combination 
$f=\sum_{i=1}^p \alpha_i f_i$  for non-zero complex coefficients
$\alpha_i$ 
 is  not BT-limited, unless 
$A_i=B_i$ for all $i=1,2,...,p$. It is easy to see
that if $A_i=B_i$ for all $i=1,2,...,p$, then the above linear combination
$f$ is indeed BT-limited and $f\in {\cal BL}_{A, B, a, b}$,
where $A=A_i$, $B=B_i$, $a=\min\{a_1,a_2,...,a_p\}$ and $b=\max\{b_1, b_2, ..., b_p\}$.

Although the above linear combination $f$ of $p$ BT-limited 
signals is generally not BT-limited, from (\ref{61}), we have
\begin{equation}\label{62}
f(t)= \sum_{i=1}^p \alpha_i  e^{j (A_i+\Omega_i)t}
\int_{-T_i}^{T_i} \frac{\sin \Omega_i (t+a_i+T_i-s)}{\pi(t+a_i+T_i-s)} q_i(s) ds, \,\, \mbox{ for } t\in (-\infty, \infty),
\end{equation}
where $\Omega_i=(B_i-A_i)/2$, $T_i=(b_i-a_i)/2$, and 
$q_i\in L^2[-T_i, T_i]$  for $i=1,2,...,p$.

\section{Conclusion}\label{sec7}
In this paper, BT-limited signal space was introduced and characterized. 
It is a subspace
of bandlimited signals where the non-zero parts of their Fourier transforms 
are also pieces of bandlimited signals. 
Some new properties about and applying 
 BT-limited signals were also presented. 
For BT-limited signals, an extrapolation from inaccurate data
with an analytic error estimate in the whole time domain exists. 
Some simulations were presented to verify the theoretical 
extrapolation results for BT-limited signals.

\begin{thebibliography}{1}


\bibitem{slepia}
D. Slepian, H. O. Pollak, and H. J. Landau, ``Prolate spheroidal wave functions I, II,'' {\em Bell Syst. Tech. J.}, vol. 44, pp. 43-84, 1961. 

\bibitem{gerch}
R. W.  Gerchberg, ``Super-resolution through error energy reduction,''
{\em  Optica Acta}, vol. 21, no. 9, pp. 709-720, 1974.

\bibitem{pap}
 A. Papoulis, 
``A new algorithm in spectral analysis and band-limited extrapolation,''
{\em IEEE Trans. Circuits Syst.}, vol. 22,  pp. 735-742, Sept. 1975. 

\bibitem{cadzow}
J. C. Cadzow, ``An extrapolation procedure for band-limited signals,''
{\em  IEEE Trans. Acoust. Speech Signal Process.}, vol.  27, pp. 4-12, Feb.
 1979.


\bibitem{xu}
W. Y. Xu and C. Chamzas, ``On the extrapolation of band-limited functions with energy constraints,''
{\em  IEEE Trans. Acoust. Speech Signal Process.}, 
vol. 31, pp. 1222-1234, Oct. 1983. 


\bibitem{sanz}
J. L. C. Sanz and T. S. Huang, 
``Some aspects of band-limited signal extrapolation: models, discrete
 approximations and noises,''
{\em  IEEE Trans. Acoust. Speech Signal Process.}, vol. 31, 
pp. 1492-1501, Dec. 1983. 


\bibitem{zhou}
X. W. Zhou and X.-G. Xia, 
``A Sanz-Huang's conjecture on band-limited signal extrapolation with noises,''
{\em IEEE Trans. Acoust. Speech Signal Process.}, vol. 37, pp. 1468-1472, 
Sept. 1989. 

\bibitem{ferr}
P. J. S. G. Ferreira, ``Noniterative and faster iterative
methods for interpolation and extrapolation,'' 
{\em IEEE Trans. Signal Process.}, 
vol. 42, , pp. 3278-3282, Nov. 1994.

\bibitem{xia}
X.-G. Xia and M. Z. Nashed, ``A method with error estimates for band-limited 
signal extrapolation from inaccurate data,''
{\em Inverse Problems}, vol. 13, pp. 1641-1661, 1997. 

\bibitem{boas}
R. P. Boas,  {\em Entire Functions}, Academic, New York, 1954.
\end{thebibliography}
\end{document}